\documentclass[a4paper,11pt]{article}
\usepackage[top=3cm,bottom=3.4cm,left=2cm,right=2cm]{geometry}

\usepackage{color}
\usepackage{bm}
\usepackage{amsmath}
\usepackage{amssymb}
\usepackage{cite}
\usepackage[dvips]{graphicx}
\DeclareFontFamily{OT1}{pzc}{}
\DeclareFontShape{OT1}{pzc}{m}{it}{<-> s * [1.10] pzcmi7t}{}
\DeclareMathAlphabet{\mathpzc}{OT1}{pzc}{m}{it}
\newcommand{\nn}{\nonumber\\}
\newcommand{\bra}[1]{\!\left<#1 \right|}
\newcommand{\ket}[1]{\left| #1 \right>\!}

\renewcommand{\thepage}{}
\makeatletter
\@addtoreset{equation}{section}
\renewcommand{\theequation}{\thesection.\@arabic\c@equation}
\makeatother
\renewcommand{\thefootnote}{\fnsymbol{footnote}}

\begin{document}
\begin{titlepage}
\title{
\vspace*{-4ex}
\hfill
\begin{minipage}{3.5cm}
\end{minipage}\\
 \bf 
The Veneziano Amplitude \\
via Mostly BRST Exact Operator
\vspace{0.5em}
}

\author{
Isao~{\sc Kishimoto},$^{1}$\footnote{\tt ikishimo@rs.socu.ac.jp}
~~~Tomoko {\sc Sasaki},$^{2}$\footnote{\tt sasaki@asuka.phys.nara-wu.ac.jp}
~~~Shigenori {\sc Seki}$^{3}$\footnote{\tt sigenori@yukawa.kyoto-u.ac.jp}
\\
\vspace{-3ex}\\
~and~ 
\\
\vspace{-3ex}\\
Tomohiko~{\sc Takahashi}$^{2}$\footnote{\tt tomo@asuka.phys.nara-wu.ac.jp}
\\
\vspace{0ex}\\
\\
$^{1}${\it
 Center for Liberal Arts and Sciences, Sanyo-Onoda City University,}\\
{\it Daigakudori 1-1-1, Sanyo-Onoda Yamaguchi 756-0884, Japan}
\vspace{1ex}
\\
$^{2}${\it Department of Physics, Nara Women's University,}\\
{\it Nara 630-8506, Japan}
\vspace{1ex}
\\
$^{3}${\it Osaka City University Advanced Mathematical Institute (OCAMI),}\\
{\it 3-3-138, Sugimoto, Sumiyoshi-ku, Osaka 558-8585, Japan}\\
\vspace{0ex}
}

\date{}
\maketitle

\begin{abstract}
\normalsize
The Veneziano amplitude is derived from fixing one degree of freedom of
$PSL(2,\mathbb{R})$ symmetry by the insertion of a mostly BRST exact
operator. 
Evaluating the five-point function which consists of four open string tachyons
and this gauge fixing operator, 
we find it equals the Veneziano amplitude up to a sign factor. 
The sign factor is interpreted as a signed intersection number.
The result implies that the mostly BRST exact
operator, which is originally used to provide two-point string amplitudes, correctly
fixes the $PSL(2,\mathbb{R})$ gauge symmetry for general amplitudes.
We conjecture an expression for general $n$-point tree amplitudes 
with an insertion of this gauge fixing operator.

\end{abstract}

\vfill\noindent
\date{17 September 2021}
\end{titlepage}

\renewcommand{\thepage}{\arabic{page}}
\renewcommand{\thefootnote}{\arabic{footnote}}
\setcounter{page}{1}
\setcounter{footnote}{0}

\tableofcontents

\section{Introduction}

Two-point open string tree amplitude is revealed to have a non-zero
result, although the amplitude is divided by the volume of the
$PSL(2,\mathbb{R})$ symmetry with three degrees of
freedom. In~\cite{Erbin:2019uiz}, it was shown that the two degrees of
freedom lead to fixing of the positions of two vertex operators of the
amplitude and the infinite volume of the residual symmetry is canceled
by the delta function for energy conservation, which is constantly
infinite due to momentum conservation and on-shell conditions. The
resulting amplitude coincides with standard free particle expression in
quantum field theories.

In the BRST formalism, a part of the present authors provided the same
result for the two-point string amplitudes in \cite{Seki:2019ycz} by
introducing the BRST invariant
operator which fixes the residual gauge symmetry.
According to \cite{Seki:2021ivm},
the operator is defined by
\begin{align}
{\mathpzc E}(z)\equiv \frac{1}{\pi}\int_{-\infty}^\infty dq\,
\Bigl(
c\partial X^0e^{-iqX^0}-i\alpha' q(\partial c)e^{-iqX^0}\Bigr).
\label{eq:B}
\end{align}
Here $X^0(z,\bar{z})$ is the string coordinate in the time direction,
$c(z)$ is the ghost field, and $z$ is a point on the boundary.
This operator can be rewritten by a mostly BRST exact expression:
\begin{align}
  {\mathpzc E}(z)= \frac{1}{\pi}\int_{-\infty}^\infty dq\,
\frac{i}{q}\, \delta_{\rm B} e^{-iqX^0}\,, 
\quad \delta_{\rm B} e^{-iqX^0} = [\,Q_{\rm B},\,e^{-iqX^0}\,] \,,
\label{eq:V0exact}
\end{align}
where $Q_{\rm B}$ is the BRST operator.
For $q\neq 0$, the integrand is indeed BRST exact, 
but it is not so at $q=0$.
Despite depending only on the time direction
and being not on-shell, the operator is Lorentz 
and conformal invariant up to BRST exact operators \cite{Seki:2019ycz}. 
The two-point amplitude is given by calculating
the three-point function in which two arbitrary on-shell open string vertex
operators and the operator $\mathpzc{E}(z)$ are inserted on the worldsheet.

It has been found that mostly BRST exact operators can be applied to
superstring theory. In \cite{Kashyap:2020tgx}, a mostly BRST exact
operator is constructed in the pure spinor formalism, by which two-point
superstring amplitudes is shown to be equal to the corresponding
expression in the field theories.

As for multipoint amplitudes, a correct three-point amplitude is derived
in \cite{Seki:2021ivm} from the four-point function with the insertions
of three on-shell vertex operators and one ${\mathpzc E}$. In this
derivation, the Feynman $i\varepsilon$ prescription plays an important
role in extracting the singularity for a momentum of ${\mathpzc E}$,
which only contributes to a physical result. In \cite{Seki:2021ivm},
two-point amplitude is also calculated by a different way in which
double of ${\mathpzc E}$ are inserted, namely one position of the
physical vertex is integrated in a four-point function.  
In this case, the $i\varepsilon$ prescription is also important to remove the
integration of the position, and we thus obtain the correct two-point
amplitude.

The $i\varepsilon$ prescription has relation to Minkowski nature of
spacetime. In Minkowski spacetime, we deal with time differently than the
spatial dimensions, and this is reflected in the representation of
${\mathpzc E}$, which includes the time direction only.  For the
$i\varepsilon$ prescription in string theories, it is significant
that the string worldsheet also should be regarded to have Lorentz
signature. Based on this principle, it is possible to introduce
$i\varepsilon$ in the representation of amplitudes in terms of the
moduli integral \cite{Witten:2013pra}.

The mostly BRST exact operator (\ref{eq:B}) is expected to fix a part of
 $PSL(2,\mathbb{R})$ symmetry correctly, as found in \cite{Seki:2019ycz}
and \cite{Seki:2021ivm}. However, it is difficult to derive general
multipoint string amplitudes explicitly from correlation functions
with ${\mathpzc E}$. 
The purpose of this paper is to derive the Veneziano
amplitude \cite{Veneziano:1968yb}, which includes one modulus, from a
five-point function with one insertion of ${\mathpzc E}$. Consequently, we
show that the mostly BRST exact operator and the $i\varepsilon$
prescription yield 
the Veneziano amplitude correctly.

We begin in subsection \ref{sec:decomposition} 
with fixing the $PSL(2,\mathbb{R})$ symmetry
that appears in the four-point open string tachyon amplitude by the use of ${\mathpzc
E}$. As a result, we obtain a five-point function represented by the two
cross-ratios, which are the same parameters of a five-point open string
amplitude \cite{Bardakci:1968rse, Virasoro:1969pd} or the five-point
case of the Koba-Nielsen amplitude \cite{Koba:1969rw}. Then we 
decompose the integration region into 12 components depending on the order
of the vertices. To introduce $i\varepsilon$ to the five-point function,
we have to understand details of the moduli space of the five-point
function. In subsection \ref{sec:singularity}, we illustrate the moduli space,
particularly its coordinates, for one of the 12 components in
reference to the results in \cite{Hanson_Sha}, where the moduli space of
the five-point function has been extensively studied to give a contour
integral expression for that. Then, we introduce $i\varepsilon$ to a
five-point function for this domain and extract the singularity
contributing to the Veneziano amplitude. For the rest of the 12
components, we have only to repeat the same analysis as that of this
domain and the details of calculations are in appendix \ref{sec:F2toF6}.
Finally, combining these results in subsection
\ref{sec:Veneziano}, 
we obtain the Veneziano amplitude by the insertion of ${\mathpzc
E}$; however, the resulting amplitude includes the sign factor depending
on the energy of external states. As discussed in \cite{Seki:2021ivm}, this
factor is interpreted by a signed intersection number associated with
the string time coordinate $X^0$. 
Section 3 is dedicated to a conclusion and
remarks on the extension to general multipoint amplitudes.

\section{The Veneziano amplitude from a five-point function}

\subsection{Decomposition of a five-point function
\label{sec:decomposition}
}

Let us consider a four open string tachyon amplitude by inserting the
vertex operators on the real
axis of the upper-half plane as the worldsheet. For simplicity, the
target space is taken as the 26-dimensional flat Minkowski spacetime.

The amplitude is defined formally by the integrals,
\begin{align}
A_4=\frac{1}{{\rm vol}PSL(2,\mathbb{R})}
\big<\int dy_1
V_1(y_1)\int dy_2 V_2(y_2)
\int dy_3 V_3(y_3)\int dy_4
V_4(y_4)\big>,
\label{eq:ampdef}
\end{align}
where $V_i$ is an open string vertex operator, $\big<\cdots \big>$
denotes a correlation function in the upper-half plane, and each integration
is over the whole real line. ${\rm vol}PSL(2,\mathbb{R})$ is the volume of the
conformal Killing group after fixing the worldsheet to the upper-half
plane.

To fix the gauge symmetry of $PSL(2,\mathbb{R})$, we fix the positions $y_3$,
$y_4$ ($y_3<y_4$) and their integrations are replaced by the ghost fields $c(y_3)$,
$c(y_4)$. 
Here one degree of freedom of $PSL(2,\mathbb{R})$ remains unfixed.  We fix this
residual symmetry by the mostly BRST exact operator inserted at $y_0$.
In the following, we suppose that the points 
$y_0$, $y_3$ and $y_4$ are in the order $y_0<y_3<y_4$, since the order 
affects only on the overall sign of the amplitude.
For this gauge fixing, the amplitude (\ref{eq:ampdef}) becomes
\begin{align}
 {\cal A}_4=ig_{\rm o}^4C_{D_2}
\bra{0}{\mathpzc E}(y_0)
\int_{-\infty}^{\infty}dy_1
V_1(y_1)
\int_{-\infty}^{\infty}dy_2
V_2(y_2)\,cV_3(y_3)\,
cV_4(y_4)\ket{0},
\end{align}
where $C_{D_2}$ is a normalization factor for disk amplitudes:
$C_{D_2}=1/(\alpha' g_{\rm o}^2)$ \cite{Polchinski:1998rq}. 
The open string coupling $g_{\rm o}$ is assigned to each vertex operator,
but not to ${\mathpzc E}$ \cite{Seki:2021ivm}.
$\ket{0}$ is the $SL(2,\mathbb{R})$ invariant
vacuum normalized as $\langle 0|0\rangle =(2\pi)^{26}\delta^{26}(0)$.
We note that $A_4$ and ${\cal A}_4$ are different in the sign as 
discussed in \cite{Seki:2021ivm}.
Using the expression (\ref{eq:V0exact}) for $\mathpzc E$,
we rewrite the amplitude as the $q$-integral
of a correlation function:
\begin{align}
 {\cal A}_4
&=ig_{\rm o}^4C_{D_2}\int_{-\infty}^\infty dq\,{\cal F}(q),
\nn
{\cal F}(q)&\equiv
\bra{0}\Big(\frac{i}{\pi q}\delta_{\rm B}e^{-iqX^{0}}(y_0)\Big)
\int_{-\infty}^{\infty}dy_1V_1(y_1)
\int_{-\infty}^{\infty}dy_2
V_2(y_2)\,cV_3(y_3)
cV_4(y_4)\ket{0}.
\label{eq:amp5}
\end{align}

For simplicity, we consider the amplitude for open string tachyons,
that is 
\begin{align}
V_i(y) \equiv e^{ip_i\cdot X}(y).
\end{align}
As in \cite{Seki:2021ivm},
we denote $e^{-iq X^0}$ in (\ref{eq:V0exact}) 
as $e^{ip_0\cdot X}$ by introducing a
covariant expression for the momentum $q$: $p_0 = (q,
0,\cdots , 0)$.  
One can calculate the correlation function ${\cal F}(q)$, 
\begin{align}
 {\cal F}(q)&=
(2\pi)^{26}\delta^{26}\big(\sum_{k=0}^4 p_k\big)
\int_{-\infty}^{\infty}dy_2\int_{-\infty}^{\infty}dy_1\,
\frac{i}{\pi q}
\Big\{
y_{03}y_{04}y_{34}
\nn
&\quad \times \left(\frac{\alpha'p_0\cdot p_1}{y_{01}}
+\frac{\alpha'p_0\cdot p_2}{y_{02}}
+\frac{\alpha'p_0\cdot p_3}{y_{03}}
+\frac{\alpha'p_0\cdot p_4}{y_{04}}
\right)
\nn
&\quad +
\frac{\alpha'}{2}p_0\cdot p_0
(y_{03}y_{34}+y_{04}y_{34})\Big\}\times \prod_{i<j}
|y_{ij}|^{2\alpha' p_i\cdot p_j},
\label{eq:F}
\end{align}
where $y_{ij}\equiv y_i-y_j$.
Moreover, it is easily found that ${\cal F}(q)$ is rewritten as
\begin{align}
{\cal F}(q)&=
(2\pi)^{26}\delta^{26}\big(\sum_{k=0}^4 p_k\big)
\int_{-\infty}^{\infty}dy_2\int_{-\infty}^{\infty}dy_1\,
\frac{-i}{2\pi q}\sum_{k=1}^2\frac{\partial}{\partial y_k}
\Big(
y_{k3}y_{k4}y_{34}\prod_{i<j}
|y_{ij}|^{2\alpha' p_i\cdot p_j}\Big).
\label{eq:BRST_WT}
\end{align}
This expression implies that ${\cal F}(q)$ is generically zero
since the integrand is given by the total derivative, 
but that is not the case for $q=0$ owing to the reciprocal factor of $q$.
It is the result of the Ward-Takahashi identity for the BRST symmetry.
Therefore, in calculating the amplitude, we have only to
evaluate the singularity of ${\cal F}(q)$ at the support $q=0$.

In order to extract the singularity at $q=0$, it is necessary to
represent the amplitude by a moduli integral\cite{Seki:2021ivm}.
Next, let us rewrite ${\cal F}(q)$ using moduli parameters, which are
that of a five-point function, although calculating a four-point open string amplitude.
We note that, for $q\neq 0$, ${\cal F}(q)$ is not invariant
under $PSL(2,\mathbb{R})$ transformation since the energy $q$ is integrated
in ${\cal F}(q)$ and there is no on-shell condition for 
the momentum of ${\mathpzc E}$.
So we will not use $PSL(2,\mathbb{R})$ symmetry to express ${\cal F}(q)$ by moduli, 
unlike the conventional case.

The moduli space of a five-point function has real dimension 2.
First, we choose the following two cross-ratios as the coordinate of
the moduli space:
\begin{align}
 u_1=\frac{y_{01}y_{34}}{y_{03}y_{14}},~~~
 u_2=\frac{y_{02}y_{34}}{y_{03}y_{24}}.
\label{eq:uv}
\end{align}
From (\ref{eq:uv}), it follows that
\begin{align}
dy_1dy_2=du_1 du_2\,\Big|\frac{y_{03}y_{14}y_{24}}{y_{04}y_{34}}
\Big|^2,
\label{eq:dy1dy2} 
\end{align}
and, by using
momentum conservation and on-shell conditions\footnote{
$\alpha'(p_k)^2=1~~~(\,k=1,\cdots,4\,)$.}, components in 
the integrand of (\ref{eq:F}) can be rewritten as
\begin{align}
&
\Big\{
y_{03}y_{04}y_{34}
\left(\frac{\alpha'p_0\cdot p_1}{y_{01}}
+\frac{\alpha'p_0\cdot p_2}{y_{02}}
+\frac{\alpha'p_0\cdot p_3}{y_{03}}
+\frac{\alpha'p_0\cdot p_4}{y_{04}}
\right)
+
\frac{\alpha'}{2}p_0\cdot p_0
(y_{03}y_{34}+y_{04}y_{34})\Big\}
\nn
&=y_{34}^2\,\Big(
\frac{\alpha'p_0\cdot p_0}{2}+\alpha'p_0\cdot p_3
+\frac{\alpha'p_0\cdot p_1}{u_1}
+\frac{\alpha' p_0\cdot p_2}{u_2}
\Big),
\label{eq:ppyij}
\end{align}
\begin{align}
\prod_{i<j}|y_{ij}|^{2\alpha'p_i\cdot p_j}
&=
\Big|\frac{y_{03}y_{14}y_{24}}{y_{04}}\Big|^{-2}
\Big|\frac{y_{03}y_{04}}{y_{34}}\Big|^{-\alpha'p_0\cdot p_0}
\nn
&\quad 
\times
|u_1|^{2\alpha'p_0\cdot p_1}|1-u_1|^{2\alpha' p_1\cdot p_3}
|u_2|^{2\alpha'p_0\cdot p_2}|1-u_2|^{2\alpha' p_2\cdot p_3}
|u_1-u_2|^{2\alpha' p_1\cdot p_2}.
\label{eq:yij}
\end{align} 
Substituting (\ref{eq:dy1dy2}), (\ref{eq:ppyij}) and (\ref{eq:yij}) into
(\ref{eq:F}), we obtain
\begin{align}
{\cal F}(q)&=
(2\pi)^{26}\delta^{26}\big(\sum_{k=0}^4 p_k\big)\,
\Big|\frac{y_{03}y_{04}}{y_{34}}\Big|^{\alpha'q^2}
\int^{\infty}_{-\infty}du_2\int^{\infty}_{-\infty}du_1\,
\frac{i}{\pi q}
\nn
&\quad 
\times
\Big(
\frac{\alpha'p_0\cdot p_0}{2}+\alpha'p_0\cdot p_3
+\frac{\alpha'p_0\cdot p_1}{u_1}
+\frac{\alpha' p_0\cdot p_2}{u_2}
\Big)
\nn
&\quad 
\times
|u_1|^{2\alpha'p_0\cdot p_1}|1-u_1|^{2\alpha' p_1\cdot p_3}
|u_2|^{2\alpha'p_0\cdot p_2}|1-u_2|^{2\alpha' p_2\cdot p_3}
|u_1-u_2|^{2\alpha' p_1\cdot p_2}.
\label{eq:Fuv}
\end{align}

The expression (\ref{eq:Fuv}) corresponds to the correlation
function which is derived from the transformation of 
the positions $y_0$, $y_1$,
$y_2$, $y_3$ and $y_4$ to $0$, $u_1$, $u_2$, $1$ and $\infty$, respectively.
This $PSL(2,\mathbb{R})$ transformation is given by
\begin{align}
 f(z)=-\frac{y_{34}(z-y_0)}{y_{03}(z-y_4)}.
\label{eq:fz}
\end{align}
It turns out that the factor with the exponent $\alpha' q^2$ is given as
$|f'(y_0)|^{\alpha' p_0\cdot p_0}$, which is the conformal factor 
of the primary field $e^{ip_0\cdot X}(y_0)$ for the $PSL(2,\mathbb{R})$ transformation
(\ref{eq:fz}).

Here, we separate the integration region of $u_1$ and $u_2$
into 12 parts, which are labeled by the cyclic order of five positions, {\it i.e.}, 
\begin{align}
& 1:~[0,1,2,3,4],\qquad
 2:~[0,1,3,2,4],\qquad
 3:~[1,2,0,3,4],
\nn
& 4:~[1,0,2,3,4],\qquad
5:~[0,3,1,2,4],\qquad
 6:~[1,0,3,2,4],
\label{eq:order}
\end{align}
and the order in which $y_1$ and $y_2$ are exchanged.  For
example, $[0,1,2,3,4]$ denotes the cyclic order $y_0\rightarrow y_1
\rightarrow y_2\rightarrow y_3\rightarrow y_4$ 
on the boundary, which corresponds to the integration region $0<u_1<u_2<1$
in the expression (\ref{eq:Fuv}).
One cyclic order corresponds to an integration region of $u_1$, $u_2$.
Therefore, ${\cal F}(q)$ can be decomposed to integrals corresponding to
each order:
\begin{align}
 {\cal F}(q)&=
(2\pi)^{26}\delta^{26}\big(\sum_{k=0}^4 p_k\big)\,
\Big|\frac{y_{03}y_{04}}{y_{34}}\Big|^{\alpha'q^2}
\Big\{
F_1(q)+F_2(q)+\cdots+F_6(q)\Big\}+(p_1\leftrightarrow p_2),
\label{eq:calF}
\end{align}
where $(p_1\leftrightarrow p_2)$ denotes the terms given by exchanging
$p_1$ and $p_2$ in the previous term. $F_i(q)~~(i=1,\cdots,6)$
is the integral over the domain corresponding to the $i$-th order
in (\ref{eq:order}):
\begin{align}
 F_1(q)&=\int^1_0du_2\int^{u_2}_0du_1\,
\frac{i}{\pi q}
\nn
&\quad 
\times
\Big(
\frac{\alpha'p_0\cdot p_0}{2}+\alpha'p_0\cdot p_3
+\frac{\alpha'p_0\cdot p_1}{u_1}
+\frac{\alpha' p_0\cdot p_2}{u_2}
\Big)
\nn
&\quad 
\times
{u_1}^{2\alpha'p_0\cdot p_1}(1-u_1)^{2\alpha' p_1\cdot p_3}
{u_2}^{2\alpha'p_0\cdot p_2}(1-u_2)^{2\alpha' p_2\cdot p_3}
(u_2-u_1)^{2\alpha' p_1\cdot p_2}
\label{eq:F1uv}
\end{align}
and so on.

\begin{figure}[htbp]
\begin{center}
\includegraphics[width=7.5cm]{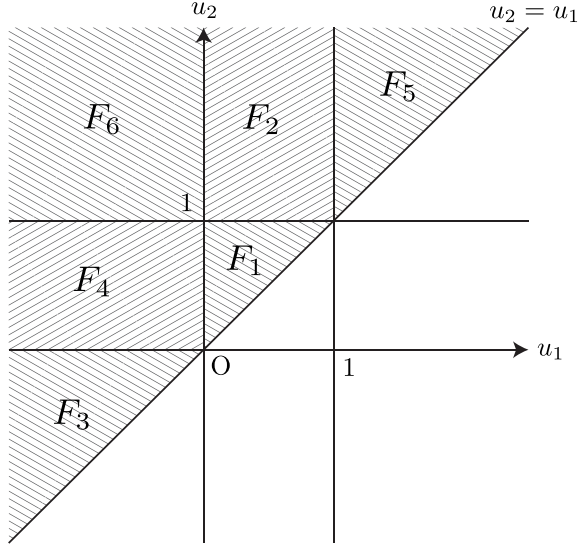} 
\end{center}
\caption{The six connected components of the domain of the parameters $(u_1,u_2)$.
Added the mark $F_i$ to show corresponding areas to the
integrals $F_i(q)$ to the figure of the 12 connected components given in
\cite{Hanson_Sha}.}
\label{fig:uv}
\end{figure}

The six components of the integration region cover the domain $u_2\geq u_1$
in the $u_1u_2$-plane.
The other six components given by the exchange of 1 and 2
fill the remaining region. These 12 components of the domain
in the $u_1u_2$-plane are the same as ones given in \cite{Hanson_Sha} to
describe the domain of parameters of five-point cross-ratios. The parameters
$u_1$, $u_2$ correspond to $s$, $t$ in \cite{Hanson_Sha}.
Here, we summarize the relationship between the integrals $F_i(q)$ and the
six components in Fig.~\ref{fig:uv}.

\subsection{Singularities for $F_i(q)$
\label{sec:singularity}
}

It is convenient to express $F_1(q)$ in terms of the different
variables $x$, $y$
given by
\begin{align}
u_1= xy,~~~u_2=x.
\label{eq:uvxy}
\end{align}
$x$, $y$ also are cross-ratios:
\begin{align}
 x=\frac{y_{02}y_{34}}{y_{03}y_{24}},~~~
 y=\frac{y_{24}y_{01}}{y_{14}y_{02}}.
\label{eq:xy}
\end{align}
The integration region of $x$, $y$ is on the unit square
$0\leq x,\,y\leq 1$ and then $F_1$ becomes
\begin{align}
 F_1(q)&=
\int^{1}_{0}dx\int^{1}_{0}dy
\,\frac{-i\alpha'}{\pi}\Big(
\frac{q}{2}
+\frac{p_1^0}{xy}
+\frac{p_2^0}{x}+p_3^0\Big)
\nn
&\quad 
\times 
x^{2\alpha'p_3\cdot p_4-\alpha'p_0\cdot p_0+1}
(1-x)^{2\alpha' p_2\cdot p_3}
y^{2\alpha'p_0\cdot p_1}
(1-y)^{2\alpha' p_1\cdot p_2}
(1-xy)^{2\alpha' p_1\cdot p_3},
\label{eq:F1xy}
\end{align}
where we have used momentum conservation and on-shell conditions.

To extract the singularity of $F_{1}(q)$ at $q=0$, 
we have to know the detail of the boundary
 of the moduli space associated with (\ref{eq:F1xy}). 
The moduli space of a five-point tree amplitude
of bosonic open strings is known as ${\cal D}_{0,5}$, which
parametrizes five cyclically ordered points $y_{0},y_{1},\cdots,y_{4}$
 on the boundary of the upper-half plane modulo $PSL(2,\mathbb{R})$ and has real
dimension two. ${\cal D}_{0,5}$ has five boundary components ${\cal
B}_{i,i+1}\ (i=0,1,\cdots,4)$\cite{Witten:2013pra}, 
which parameterize
limiting configurations arising from the coincidence of $y_i$ and
$y_{i+1}$ (we identify $y_5$ as $y_0$). 

From (\ref{eq:xy}), we find that $x=0$ leads to $y_{02}=0$ or $y_{34} = 0$. 
If $y_{02}$ becomes zero for the order $[0,1,2,3,4]$, $y_0$, $y_1$ and $y_{2}$
become coincident because $y_{1}$ is the intermediate point between $y_0$ and
 $y_2$.
This is equivalent to $y_{34}\rightarrow 0$ up to
$PSL(2,\mathbb{R})$ transformation\cite{Witten:2013pra} as illustrated as
follows: Suppose that $y_0=y_1-h,~y_2=y_1+h~(h>0)$, $y_0$, $y_1$ and
$y_2$ are mapped to $y_0'=1$, $y_1'=\infty$ and $y_2'=-1$ by the
$PSL(2,\mathbb{R})$ transformation
\begin{align}
 z'=f(z)=\frac{-h}{z-y_1}.
\end{align}
Then, we find
\begin{align}
 y_4'-y_3'=f(y_4)-f(y_3)=-h\,\frac{y_{34}}{y_{13}y_{14}}.
\label{eq:y34}
\end{align}
From (\ref{eq:y34}), as $y_0$, $y_1$ and $y_2$
becoming coincident ($h\rightarrow 0$), $y'_3$ and $y_4'$ 
are approaching each other in the mapped plane, while $y'_0$, $y'_1$ and
$y'_2$ are fixed.

As a result, in either case ($y_{02}=0$ or $y_{34}=0$), the limit 
$x\rightarrow 0$ corresponds to the boundary ${\cal B}_{34}$, and then
${\cal B}_{34}$ is parameterized by $y$.
Similarly, from (\ref{eq:xy}), we find
that ${\cal B}_{01}$ is given by the limit $y\rightarrow 0$ and
it is parameterized by $x$.

Noting that $q$ becomes zero for the non-zero result of ${\cal F}(q)$
 (\ref{eq:BRST_WT}), 
we find that in (\ref{eq:F1xy}), the limits $x\rightarrow 0$ and
$y\rightarrow 0$, namely ${\cal B}_{34}$ and ${\cal B}_{01}$, generate
poles associated with $p_3\cdot p_4$ and $p_0\cdot p_1$, respectively.
Then, it might be expected from the exponents of the integrand of
(\ref{eq:F1xy}) that the limits $x\rightarrow 1$ and $y\rightarrow 1$
describe ${\cal B}_{23}$ and ${\cal B}_{12}$. However, it is not the case,
as seen as follows. From (\ref{eq:xy}), we obtain
\begin{align}
 1-x=\frac{y_{04}y_{23}}{y_{03}y_{24}},~~~
1-y=\frac{y_{04}y_{12}}{y_{14}y_{02}}.
\end{align}
So $x=1$ implies $y_{04}=0$ or $y_{23}=0$.  In the order $[0,1,2,3,4]$,
$(y_0,y_4)$ and $(y_2,y_3)$ are pairs of adjacent points and so
$y_{04}=0$ and $y_{23}=0$ represent ${\cal B}_{40}$ and ${\cal B}_{23}$,
respectively. Then it is impossible to determine uniquely the boundary
represented by the limit $x\rightarrow 1$. Similarly, the limit
$y\rightarrow 1$ corresponds to both boundaries, ${\cal B}_{40}$ and
${\cal B}_{12}$. As a result, $(x,y)$ is not a good coordinate in the
neighborhood of the boundaries ${\cal B}_{23}$, ${\cal B}_{40}$ 
and ${\cal B}_{12}$.

To describe these boundaries, we transform $(x, y)$ to $(x',y')$ with
\begin{align}
 x'=\frac{1-x}{1-xy},~~~y'=1-xy.
\label{eq:xytrans}
\end{align}
It maps the unit square $0\leq x,\,y\leq 1$ to the unit square
$0\leq x',\,y'\leq 1$, and $(x', y')$ is given by the cross-ratios:
\begin{align}
 x'=\frac{y_{14}y_{23}}{y_{24}y_{13}},
~~~
 y'=\frac{y_{13}y_{04}}{y_{03}y_{14}}.
\label{eq:xy2}
\end{align}
According to the above discussion, from (\ref{eq:xy2}), one can easily
see that the limits $x'\rightarrow 0$ and $y'\rightarrow 0$ correspond
to ${\cal B}_{23}$ and ${\cal B}_{40}$, respectively.
For the remained boundary ${\cal B}_{12}$, we apply the
same transformation as (\ref{eq:xytrans}) for
$(x', y')$:
\begin{align}
 x''=\frac{1-x'}{1-x'y'},
~~~y''=1-x'y'.
\label{eq:xytrans2}
\end{align}
Again, $(x'', y'')$ is a point in the unit square.
$x''$ is also the cross-ratio, 
\begin{align}
 x''=\frac{y_{03}y_{12}}{y_{02}y_{13}},
\label{eq:x2}
\end{align}
and we find $y''=x$.  From (\ref{eq:x2}), it is clear
that ${\cal B}_{12}$ is given by the limit $x''\rightarrow 0$.

\begin{figure}[htbp]
\begin{center}
\includegraphics[width=6cm]{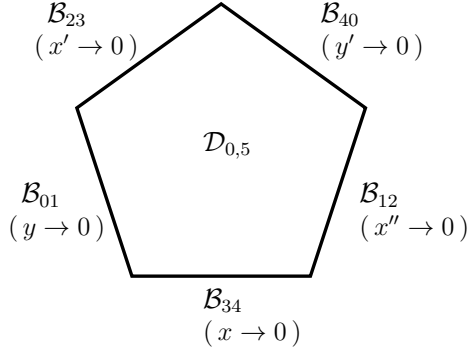} 
\end{center}
\caption{The moduli space of five-point open string amplitudes
corresponding to the order $[0,1,2,3,4]$. Added the limits describing
the boundaries to the pentagon depicted in \cite{Witten:2013pra}.}
\label{fig:D05}
\end{figure}

By using (\ref{eq:xytrans}) and (\ref{eq:xytrans2}), we have the five
cross-ratios, $x$, $y$, $x'$, $y'$ and $x''$.  These correspond to
five-point cross-ratios used in \cite{Hanson_Sha} to construct the
contour integral representation for the dual five-point function.  As in
\cite{Witten:2013pra}, the moduli space ${\cal D}_{0,5}$ can be depicted
as a pentagon, and its five boundary components correspond to ${\cal
B}_{i,i+1}$.  Finally, we summarize these results about the
boundaries in Fig.~\ref{fig:D05}.

\begin{figure}[htbp]
\begin{center}
\includegraphics[width=7cm]{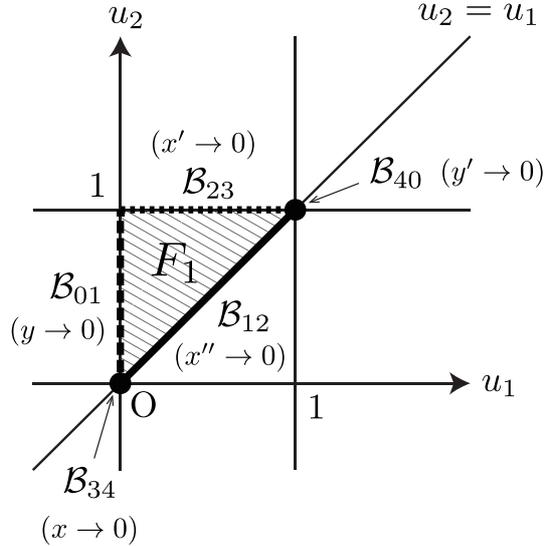}
\end{center}
\caption{The correspondence of the integration domain and the moduli space
for $F_1(q)$.
\label{fig:F1}}
\end{figure}

Let us clarify the correspondence between the pentagon and the triangle
for $F_1(q)$. By using (\ref{eq:uvxy}), (\ref{eq:xytrans}) and
(\ref{eq:xytrans2}), $(u_{1},u_{2})$ can be written by the five-point cross-ratios:
\begin{align}
&(u_1,\,u_2)=(xy,\,x)=(1-y^{\prime},\,1-x^{\prime}y^{\prime})=
\left(
\frac{y^{\prime\prime}
(1-x^{\prime\prime})}{1-x^{\prime\prime}y^{\prime\prime}},\,
y^{\prime\prime}
\right).
\label{eq:uvxy2}
\end{align}
Therefore, we find the relation between the boundaries ${\cal B}_{i,i+1}$
and parts of the triangle. As shown in Fig.~\ref{fig:F1}, 
the boundaries ${\cal B}_{34}$
and ${\cal B}_{40}$ shrink to the points: $(0,0)$
and $(1,1)$ in the $u_1u_2$-plane.
By using (\ref{eq:uvxy2}), we change the variables in (\ref{eq:F1xy}),
and then $F_1(q)$ is expressed in terms of $x'$ and $y'$ as
\begin{align}
 F_1(q)
&=
\int_0^1dx'\int_0^1dy' \frac{-i\alpha'}{\pi}\,
\Big(\frac{q}{2}+p_3^0
+p_1^0\frac{1}{1-y'}
+p_2^0\frac{1}{1-x'y'}
\Big)
\nn
&\quad
\times {x'}^{2\alpha'p_2\cdot p_3}
(1-x')^{2\alpha'p_1\cdot p_2}
\nn
&\quad
\times
{y'}^{2\alpha'p_0\cdot p_4+\alpha'p_0\cdot p_0-1}
(1-y')^{2\alpha'p_0\cdot p_1}
(1-x'y')^{2\alpha'p_0\cdot p_2},
\label{eq:F1xy2}
\end{align}
and in terms of $x''$ and $y''$ as
\begin{align}
 F_1(q)
&=
\int_0^1dx''\int_0^1dy'' \frac{-i\alpha'}{\pi}
\Big(
\frac{q}{2}+p_3^0
+p_1^0\frac{1-x''y''}{(1-x'')y''}
+p_2^0\frac{1}{y''}
\Big)
\nn
&\quad
\times {x''}^{2\alpha'p_1\cdot p_2}
(1-x'')^{2\alpha'p_0\cdot p_1}
\nn
&\quad
\times
{y''}^{2\alpha'p_3\cdot p_4-\alpha'p_0\cdot p_0+1}
(1-y'')^{2\alpha'p_0\cdot p_4+\alpha'p_0\cdot p_0-1}
(1-x''y'')^{2\alpha'p_1\cdot p_4}.
\label{eq:F1xy3}
\end{align}

Now that the coordinates describing the neighborhood of
all boundaries of ${\cal D}_{0,5}$ are given,
let us evaluate the singularity of
$F_1(q)$ at $q=0$.  From (\ref{eq:F1xy}), (\ref{eq:F1xy2}) and
(\ref{eq:F1xy3}), it turns out that the boundaries associated with $y_0$,
which are ${\cal B}_{01}$ and ${\cal B}_{40}$,
generate the delta function with respect to $q$.
For ${\cal B}_{01}$, we expand the integrand of (\ref{eq:F1xy})
around $y=0$ as follows:
\begin{align}
 F_1(q)&=
\int^{1}_{0}dx\int^{1}_{0}dy
\,\frac{-i\alpha'}{\pi} \,
y^{2\alpha'p_0\cdot p_1-1}
\Big\{p_1^0\,x^{2\alpha'p_3\cdot p_4-\alpha'p_0\cdot p_0}
(1-x)^{2\alpha' p_2\cdot p_3}+{\mathrm O}(y)\Big\}.
\end{align}
By the Feynman $i\varepsilon$ prescription, we extract the singularity
given by a neighborhood of $y=0$ \cite{Witten:2013pra}\footnote{See also (6.4.12) in \cite{Polchinski:1998rq}.}:
\begin{align}
 \int^{1}_{0}dy
y^{2\alpha'p_0\cdot p_1-1}
\sim  \frac{1}{-2\alpha' q p_1^0-i\varepsilon}
\sim  \frac{\pi i}{2\alpha'|p_1^0|}\delta(q),
\end{align}
where $\sim$ stands for equal up to terms for $q\neq 0$, which should be
vanished by the BRST symmetry as seen in (\ref{eq:BRST_WT}).
Similarly, we can evaluate the singularity associated with ${\cal B}_{40}$
($y^{\prime}\to 0$) in terms of (\ref{eq:F1xy2}). Finally, the singularity of $F_1(q)$ is given
by
\begin{align}
 F_1(q) &\sim \frac{1}{2}\left(\frac{p_1^0}{|p_1^0|}
-\frac{p_4^0}{|p_4^0|}\right)\delta(q)I(u,s),
\label{eq:F1delta}
\end{align}
where we introduce the Mandelstam variables as
\begin{align}
 s=-(p_1+p_2)^2,~~~
 t=-(p_1+p_3)^2,~~~
 u=-(p_1+p_4)^2,
\end{align}
and $I(u,s)$ is defined by using the Euler beta function $B(a,b)$:
\begin{align}
 I(u,s)&\equiv B(-\alpha'u-1,\,-\alpha' s-1),
\nn
 B(a,b)&= \int_0^1 dx\,x^{a-1}(1-x)^{b-1}.
\end{align}

To evaluate the singularity at $q=0$ for other $F_i(q)\ (i=2,\cdots,6)$,
in a similar way for $F_1(q)$, we
change the variables $u_1$, $u_2$ by appropriate coordinates
describing the boundary of the moduli space.
The details of the calculation are shown in
appendix A.  Finally, we obtain the singularities of $F_i(q)$, 
\begin{align}
 F_2(q) 
&\sim \frac{1}{2}\left(\frac{p_1^0}{|p_1^0|}
-\frac{p_4^0}{|p_4^0|}\right)\delta(q)I(t,u),
\label{eq:F2delta}
\end{align}
\begin{align}
 F_3(q) 
&\sim \frac{1}{2}\left(\frac{p_3^0}{|p_3^0|}
-\frac{p_2^0}{|p_2^0|}\right)\delta(q)I(u,s),
\label{eq:F3delta}
\end{align}
\begin{align}
 F_4(q) 
&\sim \frac{1}{2}\left(\frac{p_2^0}{|p_2^0|}
-\frac{p_1^0}{|p_1^0|}\right)\delta(q)I(u,s),
\label{eq:F4delta}
\end{align}
\begin{align}
 F_5(q) 
&\sim \frac{1}{2}\left(\frac{p_3^0}{|p_3^0|}
-\frac{p_4^0}{|p_4^0|}\right)\delta(q)I(s,t),
\label{eq:F5delta}
\end{align}
\begin{align}
 F_6(q) 
&\sim \frac{1}{2}\left(\frac{p_3^0}{|p_3^0|}
-\frac{p_1^0}{|p_1^0|}\right)\delta(q)I(t,u).
\label{eq:F6delta}
\end{align}

\subsection{Veneziano amplitude
\label{sec:Veneziano}
}

Now we have evaluated the singularities at $q=0$ of all $F_i(q)$, namely
those of ${\cal F}(q)$ given by (\ref{eq:calF}).  Noting that ${\cal
F}(q)\ (q\neq 0)$ has no contribution to the amplitude as mentioned in
(\ref{eq:BRST_WT}), we can calculate the amplitude (\ref{eq:amp5}) by
using (\ref{eq:calF}) and by adding together (\ref{eq:F1delta}),
(\ref{eq:F2delta}), (\ref{eq:F3delta}), (\ref{eq:F4delta}),
(\ref{eq:F5delta}) and (\ref{eq:F6delta}).  The resulting amplitude
 is\footnote{Note that $p_1\leftrightarrow p_2$ induces
$t\leftrightarrow u$.}
\begin{align}
 {\cal A}_4&
=\frac{1}{2}\Big(\frac{p_3^0}{|p_3^0|}
-\frac{p_4^0}{|p_4^0|}
\Big)\times A_4,
\label{eq:amp_result}
\end{align}
where $A_4$ is the Veneziano amplitude, which is same as (6.4.9)
in~\cite{Polchinski:1998rq}:
\begin{align}
A_4=2ig_{\rm o}^4 C_{D_2}(2\pi)^{26}\delta^{26}\big(\sum_{k=1}^4 p_k\big)\,
(I(s,t)+I(t,u)+I(u,s)).
\end{align}
Thus, we have derived the Veneziano amplitude from the five-point function where
one of the vertices is the mostly BRST exact operator.

We should comment that the sign factor in (\ref{eq:amp_result}) is
interpreted as the signed intersection number as discussed
in~\cite{Seki:2021ivm}. The mostly BRST exact operator is constructed by
the gauge fixing condition $X^0(y_0)=0$, and then the Faddev-Popov
determinant for the gauge fixing includes $\partial X^0(y_0)$. The
important point is that we do not take the absolute value of the
determinant as a conventional operator formalism. Moreover, the fixed
operators at $y_3$ and $y_4$ correspond to asymptotic states in the
scattering process, and so the time coordinates $X^0(y_3)$ and $X^0(y_4)$
become $-\infty$ or $+\infty$ depending on whether the state is incoming
or outgoing. Consequently, the sign of the amplitude is given as the signed
intersection number of the graph $u=X^0(y)$ with $u=0$, and this is
determined by whether the fixed vertices are incoming or outgoing. It is
easily seen that the sign does not depend on the momenta of the
integrated vertices.  As a result, the sign factor of
(\ref{eq:amp_result}) is in agreement with that of the three-point amplitude
in \cite{Seki:2021ivm} derived from the four-point function with ${\mathpzc E}$.

\section{Concluding remarks
\label{sec:remarks}}

We have shown that the five-point function for four open string tachyon
vertices and the mostly BRST exact operator ${\mathpzc E}$
coincides with the Veneziano
amplitude up to the sign factor, which depends on the external momenta
and takes the value, $\pm 1$ or 0.
In the calculation, the contribution from the singularity near the 
boundary of the moduli space was significant.
We have used the Feynman $i\varepsilon$ prescription to extract the
singularities after \cite{Seki:2021ivm}.

The result of amplitudes with the insertion of ${\mathpzc E}$ is
represented by the following equation:
\begin{align}
&
ig_{\rm o}^nC_{D_2}
\bra{0}{\mathpzc E}(y_0)
\int_{-\infty}^{\infty}dy_1
V_1(y_1)\cdots 
\int_{-\infty}^{\infty}dy_{n-2}
V_{n-2}(y_{n-2})\,cV_{n-1}(y_{n-1})\,
cV_n(y_n)\ket{0}
\nn
&=
\frac{1}{2}\Big(\frac{p_{n-1}^0}{|p_{n-1}^0|}
-\frac{p_n^0}{|p_n^0|}
\Big)\times A_n,
\label{eq:AnAn}
\end{align}
where $A_n$ is the correct $n$-point open string amplitude for the
vertices $V_i(y_i)\ (i=1,\cdots,n)$.
For $n=2$, this is shown in
\cite{Seki:2019ycz} as two-point amplitudes for arbitrary vertices.  For
$n=3$, this is given in \cite{Seki:2021ivm} for tachyon vertex
operators. Then, in the case of $n=4$ for tachyon vertices, $A_n$ is the
Veneziano amplitude, and this is proven in this paper.  For tachyon
vertices, $A_n\ (n\geq 5)$ is given by the Koba-Nielsen amplitude 
and the sign factor is expected to be the same as $n=2,3,4$ 
from the interpretation of the signed intersection number. 
The equation \eqref{eq:AnAn} is conjectured to hold in this general case $(n\geq 5)$. 

We expect that similar identities hold for arbitrary vertex operators.
In fact, for $n=2$, (\ref{eq:AnAn}) is true for arbitrary external
states\cite{Seki:2019ycz, Seki:2021ivm}. To prove these, it is
necessary to understand the moduli space for corresponding amplitudes
similarly to
the derivation of the Veneziano amplitude from the five-point function.

\section*{Acknowledgments}
This work was supported in part by JSPS
Grant-in-Aid for Scientific Research (C) \#20K03972.
I.~K.~was supported in part by JSPS
Grant-in-Aid for Scientific Research (C) \#20K03933.
S.~S.~was supported in part by MEXT Joint
Usage/Research Center on Mathematics and Theoretical Physics at OCAMI 
and by JSPS Grant-in-Aid for Scientific Research (C) \#17K05421.
\appendix

\section{Evaluation of $F_i(q)~~~(i=2,\cdots,6)$
\label{sec:F2toF6}}

\subsection{$F_2(q)$}

\begin{figure}[htbp]
\begin{center}
\includegraphics[width=7cm]{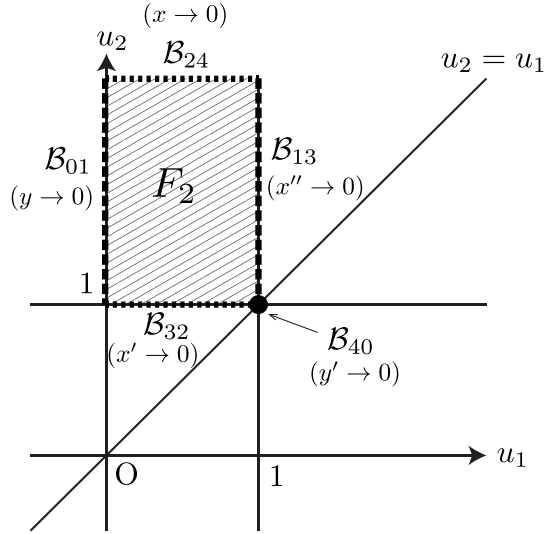}
\end{center}
\caption{The correspondence of the integration domain and the moduli space
for $F_2(q)$.
\label{fig:F2}}
\end{figure}

According to Fig.~\ref{fig:uv}, the integral
$F_2(q)$ is given by the form
\begin{align}
 F_2(q)&= \int_1^\infty du_2 \int_0^1 du_1 \cdots.
\end{align}
By changing the variables, $u_1=y$, $u_2=1/x$, we can express $F_2(q)$ as
the integral over the unit square:
\begin{align}
 F_2(q)
&=
\int_0^1dx\int_0^1dy \frac{-i\alpha'}{\pi}
\Big(\frac{q}{2}+p^0_3
+p_1^0\frac{1}{y}+p_2^0 \,x\Big)
\nn
&\quad
\times x^{2\alpha'p_2\cdot p_4}
(1-x)^{2\alpha'p_2\cdot p_3}
y^{2\alpha'p_0\cdot p_1}
(1-y)^{2\alpha'p_1\cdot p_3}
(1-xy)^{2\alpha'p_1\cdot p_2}.
\label{eq:F2xy}
\end{align}
By the transformations (\ref{eq:xytrans}) and (\ref{eq:xytrans2}) for
$(x,y)$, we can obtain different expressions for $F_2(q)$:
\begin{align}
 F_2(q)
&=
\int_0^1dx'\int_0^1dy' \frac{-i\alpha'}{\pi}
\Big(\frac{q}{2}+p^0_3
+p^0_1\,\frac{1-x'y'}{1-y'}
+p^0_2\,(1-x'y')
\Big)
\nn
&\quad
\times {x'}^{2\alpha'p_2\cdot p_3}
(1-x')^{2\alpha'p_1\cdot p_3}
{y'}^{2\alpha'p_0\cdot p_4+\alpha'p_0\cdot p_0-1}
(1-y')^{2\alpha'p_0\cdot p_1}
(1-x'y')^{2\alpha'p_0\cdot p_3+\alpha'p_0\cdot p_0-1}
\label{eq:F2xy2}
\\
&=
\int_0^1dx''\int_0^1dy'' \frac{-i\alpha'}{\pi}
\Big(\frac{q}{2}+p^0_3+p^0_1\,\frac{1-x''y''}{1-x''}
+p^0_2\,y''
\Big)
\nn
&\quad
\times {x''}^{2\alpha'p_1\cdot p_3}
(1-x'')^{2\alpha'p_1\cdot p_0}
{y''}^{2\alpha'p_2\cdot p_4}
(1-y'')^{2\alpha'p_0\cdot p_4+\alpha'p_0\cdot p_0-1}
(1-x''y'')^{2\alpha'p_1\cdot p_4}.
\end{align}
Since $u_1$ and $u_2$ are given by (\ref{eq:uv}),
the five-point cross-ratios for $F_2(q)$ are explicitly written as 
\begin{align}
&
x=\dfrac{y_{03}y_{24}}{y_{02}y_{34}},\quad
y=\dfrac{y_{01}y_{34}}{y_{03}y_{14}},\quad
x^{\prime}=-\dfrac{y_{14}y_{23}}{y_{12}y_{34}},\quad
y^{\prime}=\dfrac{y_{04}y_{12}}{y_{02}y_{14}},\quad
x^{\prime\prime}=\dfrac{y_{02}y_{13}}{y_{03}y_{12}}.
\end{align}
From these expressions, we determine the limits corresponding to
the boundaries of the moduli space for the order $[0,1,3,2,4]$:
\begin{align}
&{\cal B}_{24}:(x\to 0),
&&{\cal B}_{01}:(y\to 0),
&&{\cal B}_{32}:(x^{\prime}\to 0),
&&{\cal B}_{40}:(y^{\prime}\to 0),
&&{\cal B}_{13}:(x^{\prime\prime}\to 0).
\end{align}
The original parameters $u_1$ and $u_2$ are written using
the five-point cross-ratios such as
\begin{align}
&(u_1,\,u_2)
=\Big(y,\,\frac{1}{x}\Big)
=\left(\frac{1-y^{\prime}}{1-x^{\prime}y^{\prime}},
\frac{1}{1-x^{\prime}y^{\prime}}\right)=
\left(
\frac{1-x^{\prime\prime}}{1-x^{\prime\prime}y^{\prime\prime}},
\frac{1}{y^{\prime\prime}}
\right).
\end{align}
Therefore we depict the domain for $F_2(q)$ and the boundaries of
the moduli space as in Fig.~\ref{fig:F2}. We find that ${\cal B}_{40}$
shrinks to a point in the $u_1u_2$-plane.

Now let us evaluate the singularity of $F_2(q)$, which arises from the
boundaries ${\cal B}_{01}$ and ${\cal B}_{40}$.  Similar to the case of
$F_1(q)$, we expand the integrand of (\ref{eq:F2xy}) around $y=0$ for
${\cal B}_{01}$ and that of (\ref{eq:F2xy2}) around $y'=0$ for ${\cal
B}_{40}$. Then we extract the delta function of $q$ by using
the $i\varepsilon$ prescription.  We obtain the resulting
expression for $F_2(q)$ in (\ref{eq:F2delta}).

\subsection{$F_3(q)$}

\begin{figure}[htbp]
\begin{center}
\includegraphics[width=7cm]{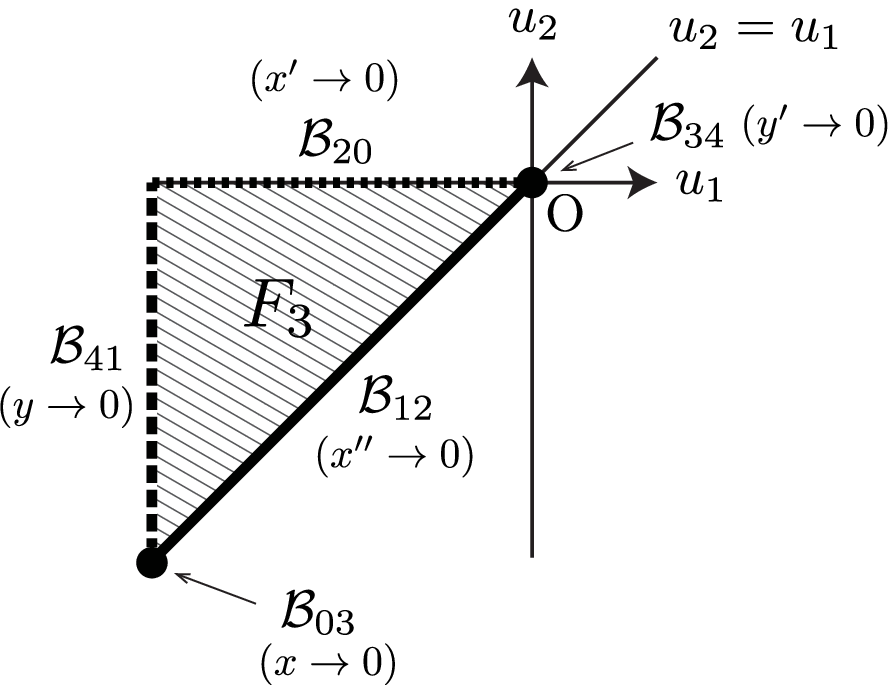}
\end{center}
\caption{The correspondence of the integration domain and the moduli space
for $F_3(q)$.
\label{fig:F3}}
\end{figure}

For the order $[1,2,0,3,4]$, from Fig.~\ref{fig:uv},
it turns out that the integral $F_3(q)$ is written as
\begin{align}
 F_3(q)&=\int_{-\infty}^0 du_2\int_{-\infty}^{u_2} du_1\cdots.
\end{align}
We change the variables by a $PSL(2,\mathbb{R})$ transformation:
\begin{align}
 u'_1=\frac{1}{1-u_1},\quad 
 u'_2=\frac{1}{1-u_2},
\end{align}
to map
the integration region onto the triangle domain as $F_1(q)$.
Then $F_3(q)$ is expressed as
\begin{align}
 F_3(q)&= \int_0^1 du_2' \int_0^{u_2'} du_1'\frac{1}{(u_1'u_2')^2}
\cdots.
\end{align}
Since the integration range of $F_3(q)$ becomes equal to that of $F_1(q)$,
we follow the same procedure applied to evaluate the singularity of $F_1(q)$.
Using $u_1'=xy$ and $u_2'=x$, we change the coordinates 
by (\ref{eq:xytrans}) and (\ref{eq:xytrans2}), 
and we have the five-point cross-ratios: 
\begin{align}
&x=\dfrac{y_{03}y_{24}}{y_{23}y_{04}},
&&y=\dfrac{y_{23}y_{14}}{y_{13}y_{24}},
&&x^{\prime}=\dfrac{y_{13}y_{02}}{y_{01}y_{23}},
&&y^{\prime}=-\dfrac{y_{01}y_{34}}{y_{04}y_{13}},
&&x^{\prime\prime}=-\dfrac{y_{12}y_{04}}{y_{01}y_{24}}.
\end{align}
The boundaries of the moduli space are given by the limits:
\begin{align}
&{\cal B}_{03}:(x\to 0),
&&{\cal B}_{41}:(y\to 0),
&&{\cal B}_{20}:(x^{\prime}\to 0),
&&{\cal B}_{34}:(y^{\prime}\to 0),
&&{\cal B}_{12}:(x^{\prime\prime}\to 0).
\end{align}
$u_1$ and $u_2$ are expressed by the five-point cross-ratios as
\begin{align}
&(u_1,\,u_2)=\left(\frac{xy-1}{xy},\,\frac{x-1}{x}\right)
=\left(\frac{y^{\prime}}{y^{\prime}-1},\,
\frac{x^{\prime}y^{\prime}}{x^{\prime}y^{\prime}-1}\right)
=\left(
\frac{y^{\prime\prime}-1}{y^{\prime\prime}(1-x^{\prime\prime})},\,
\frac{y^{\prime\prime}-1}{y^{\prime\prime}}
\right).
\end{align}
So we summarize the domain for $F_3(q)$ in the $u_{1}u_{2}$-plane in Fig.~\ref{fig:F3}.

By using the five-point cross-ratios, $F_3(q)$ can be written as
\begin{align}
 F_3(q)
&=
\int_0^1dx\int_0^1dy \frac{-i\alpha'}{\pi}
\Big(\frac{q}{2}+p^0_3
+
p_1^0\,\frac{xy}{xy-1}+p^0_2
\frac{x}{x-1}\Big)
\nn
&\quad
\times x^{2\alpha'p_0\cdot p_3+\alpha' p_0\cdot p_0-1}
(1-x)^{2\alpha'p_0\cdot p_2}
y^{2\alpha'p_1\cdot p_4}
(1-y)^{2\alpha'p_1\cdot p_2}
(1-xy)^{2\alpha'p_0\cdot p_1}
\label{eq:F3xy1}
\\
&=
\int_0^1dx'\int_0^1dy' \frac{-i\alpha'}{\pi}
\Big(\frac{q}{2}+p^0_3
+p^0_1\,\frac{y'-1}{y'}
+p^0_2\frac{x'y'-1}{x'y'}
\Big)
\nn
&\quad
\times {x'}^{2\alpha'p_0\cdot p_2}
(1-x')^{2\alpha'p_1\cdot p_2}
{y'}^{2\alpha'p_3\cdot p_4-\alpha'p_0\cdot p_0+1}
(1-y')^{2\alpha'p_1\cdot p_4}
(1-x'y')^{2\alpha'p_2\cdot p_4}
\label{eq:F3xy2}
\\
&=
\int_0^1dx''\int_0^1dy'' \frac{-i\alpha'}{\pi}
\Big(\frac{q}{2}+p^0_3
+p^0_1\,\frac{(1-x'')y''}{y''-1}
+p^0_2\,\frac{y''}{y''-1}
\Big)
\nn
&\quad
\times {x''}^{2\alpha'p_1\cdot p_2}
(1-x'')^{2\alpha'p_1\cdot p_4}
{y''}^{2\alpha'p_0\cdot p_3+\alpha'p_0\cdot p_0-1}
(1-y'')^{2\alpha'p_3\cdot p_4-\alpha'p_0\cdot p_0+1}
(1-x''y'')^{2\alpha'p_1\cdot p_3}.
\label{eq:F3xy3}
\nn
\end{align}
Since the boundaries ${\cal B}_{03}$ and ${\cal B}_{20}$ contribute to
the singularity, we have only to expand (\ref{eq:F3xy1})
and (\ref{eq:F3xy2}) around $x=0$ and $x'=0$, respectively.
We obtain the final result for $F_3(q)$ as (\ref{eq:F3delta}).

\subsection{$F_4(q)$}

\begin{figure}[htbp]
\begin{center}
\includegraphics[width=7cm]{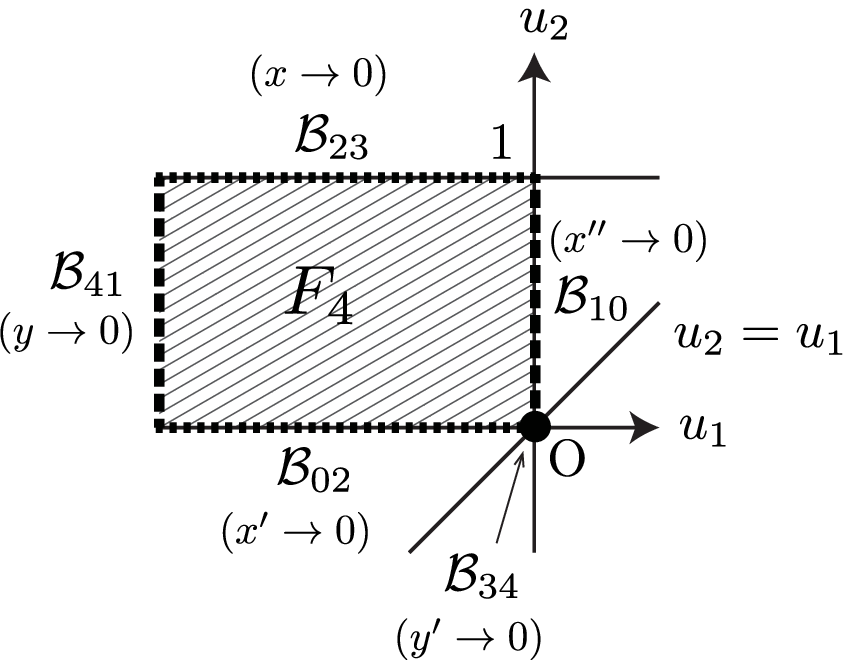}
\end{center}
\caption{The correspondence of the integration domain and the moduli space
for $F_4(q)$.
\label{fig:F4}}
\end{figure}

The integration region of $F_4(q)$ is depicted in Fig.~\ref{fig:uv}.
Similar to the case of $F_{2}(q)$, 
we can map this region onto the unit square by
\begin{align}
x=1-u_{1},\qquad y=\frac{1}{1-u_{2}}.
\label{eq:uF4_1}
\end{align}
After that, we follow the same procedure as for $F_1$, $F_2$ and $F_3$.
We can introduce the five-point cross-ratios using the
transformations (\ref{eq:xytrans}) and (\ref{eq:xytrans2}):
\begin{align}
&x=\dfrac{y_{23}y_{04}}{y_{03}y_{24}},
&&y=\dfrac{y_{03}y_{14}}{y_{13}y_{04}},
&&x^{\prime}=\dfrac{y_{02}y_{13}}{y_{03}y_{12}},
&&y^{\prime}=\dfrac{y_{12}y_{34}}{y_{24}y_{13}},
&&x^{\prime\prime}=-\dfrac{y_{01}y_{24}}{y_{04}y_{12}}.
\end{align}
So the five boundaries of the moduli space are given by the limits:
\begin{align}
&{\cal B}_{23}:(x\to 0),
&&{\cal B}_{41}:(y\to 0),
&&{\cal B}_{02}:(x^{\prime}\to 0),
&&{\cal B}_{34}:(y^{\prime}\to 0),
&&{\cal B}_{10}:(x^{\prime\prime}\to 0).
\end{align}
Then $(u_1,u_2)$ is written by
\begin{align}
&(u_1,\,u_2)
=\left(\frac{y-1}{y},1-x\right)
=\left(\frac{y^{\prime}(1-x^{\prime})}{y^{\prime}-1},\,
x^{\prime}y^{\prime}\right)
=\left(
\frac{x^{\prime\prime}(1-y^{\prime\prime})}{x^{\prime\prime}-1},\,
1-y^{\prime\prime}
\right).
\label{eq:uF4_2}
\end{align}
The domain for $F_4(q)$ and the boundaries in the $u_{1}u_{2}$-plane 
are shown in Fig.~\ref{fig:F4}.

By the transformations (\ref{eq:uF4_2}),
$F_4(q)$ can be expressed as
\begin{align}
 F_4(q)
&=
\int_0^1dx\int_0^1dy \frac{-i\alpha}{\pi}
\Big(\frac{q}{2}+p^0_3
+p^0_1\,\frac{y}{y-1}
+p^0_2\frac{1}{1-x}
\Big)
\nn
&\quad
\times x^{2\alpha'p_2\cdot p_3}
(1-x)^{2\alpha'p_0\cdot p_2}
y^{2\alpha'p_1\cdot p_4}
(1-y)^{2\alpha'p_0\cdot p_1}
(1-xy)^{2\alpha'p_1\cdot p_2}
\label{eq:F4xy1}
\\
&=
\int_0^1dx'\int_0^1dy' \frac{-i\alpha'}{\pi}
\Big(\frac{q}{2}+p^0_3
+p^0_1\frac{y'-1}{(1-x')y'}
+p^0_2\frac{1}{x'y'}
\Big)
\nn
&\quad
\times {x'}^{2\alpha'p_0\cdot p_2}
(1-x')^{2\alpha'p_0\cdot p_1}
{y'}^{2\alpha'p_3\cdot p_4-\alpha'p_0\cdot p_0+1}
(1-y')^{2\alpha'p_1\cdot p_4}
(1-x'y')^{2\alpha'p_0\cdot p_4+\alpha'p_0\cdot p_0-1}
\\
&=
\int_0^1dx''\int_0^1dy'' \frac{-i\alpha'}{\pi}
\Big(\frac{q}{2}+p^0_3
+p^0_1\frac{x''-1}{x''(1-y'')}
+p^0_2\frac{1}{1-y''}
\Big)
\nn
&\quad
\times {x''}^{2\alpha'p_0\cdot p_1}
(1-x'')^{2\alpha'p_1\cdot p_4}
{y''}^{2\alpha'p_2\cdot p_3}
(1-y'')^{2\alpha'p_3\cdot p_4-\alpha'p_0\cdot p_0+1}
(1-x''y'')^{2\alpha'p_1\cdot p_3}.
\label{eq:F4xy2}
\end{align}
To extract the singularity, we expand the integrand of $F_4(q)$ around
$x'=0$ for ${\cal B}_{02}$ and $x''=0$ for ${\cal B}_{10}$.
The resulting expression for $F_4(q)$ becomes (\ref{eq:F4delta}).

\subsection{$F_5(q)$}

\begin{figure}[htbp]
\begin{center}
\includegraphics[width=7cm]{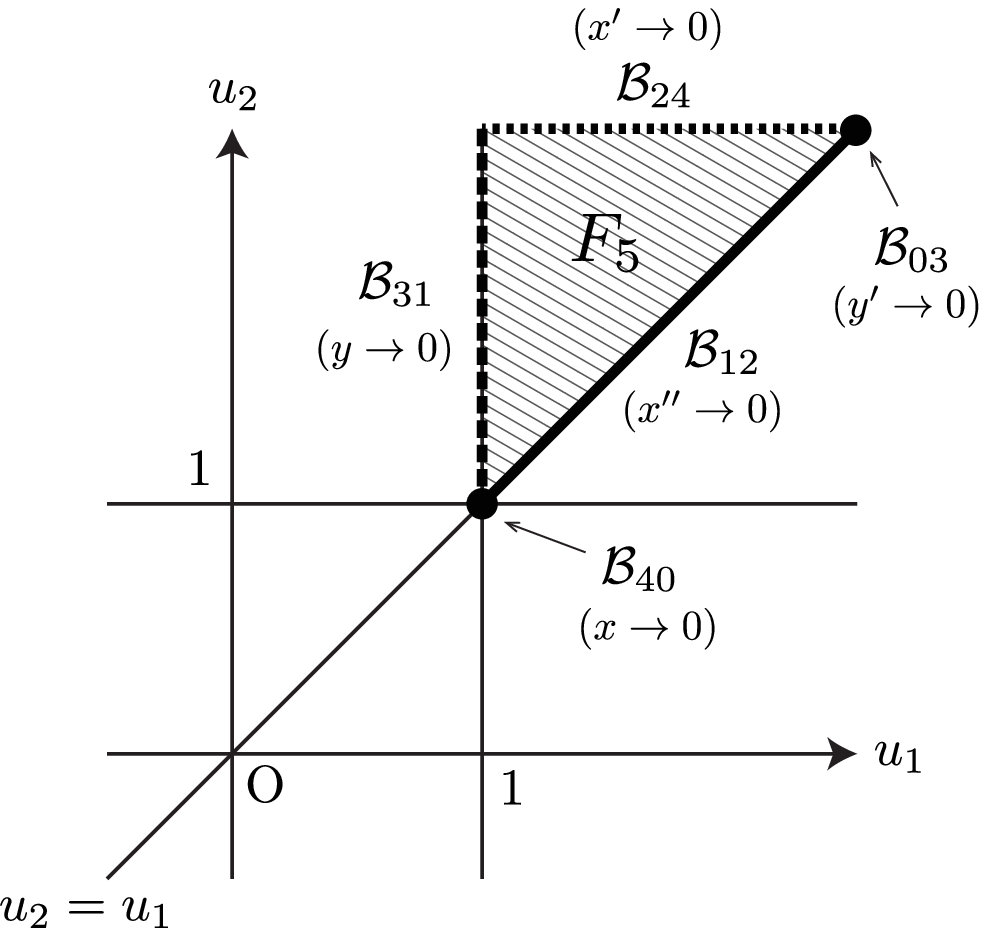}
\end{center}
\caption{The correspondence of the integration domain and the moduli space
for $F_5(q)$.
\label{fig:F5}}
\end{figure}

As for $F_5(q)$, we can map the integration region in Fig.~\ref{fig:uv}
 onto the triangle for $F_1(q)$ by 
 a $PSL(2,\mathbb{R})$ transformation: 
\begin{align}
u_{1}^{\prime\prime}=\frac{u_{1}-1}{u_{1}},\qquad
u_{2}^{\prime\prime}=\frac{u_{2}-1}{u_{2}}.
\end{align}
With $u_{1}^{\prime\prime}=xy$ and $u_{2}^{\prime\prime}=x$, 
the five-point cross-ratios are introduced by (\ref{eq:xytrans}) and (\ref{eq:xytrans2}):
\begin{align}
&x=-\dfrac{y_{04}y_{23}}{y_{02}y_{34}},
&&y=\dfrac{y_{02}y_{13}}{y_{01}y_{23}},
&&x^{\prime}=\dfrac{y_{01}y_{24}}{y_{02}y_{14}},
&&y^{\prime}=\dfrac{y_{03}y_{14}}{y_{01}y_{34}},
&&x^{\prime\prime}=-\dfrac{y_{12}y_{34}}{y_{23}y_{14}}.
\end{align}
Therefore, the boundaries of the moduli space are given by
\begin{align}
&{\cal B}_{40}:(x\to 0),
&&{\cal B}_{31}:(y\to 0),
&&{\cal B}_{24}:(x^{\prime}\to 0),
&&{\cal B}_{03}:(y^{\prime}\to 0),
&&{\cal B}_{12}:(x^{\prime\prime}\to 0).
\end{align}
$(u_{1},u_{2})$ is expressed as
\begin{align}
&(u_1,\,u_2)=\left(\frac{1}{1-xy},\frac{1}{1-x}\right)
=\left(\frac{1}{y^{\prime}},\,\frac{1}{x^{\prime}y^{\prime}}\right)
=\left(
\frac{1-x^{\prime\prime}y^{\prime\prime}}{1-y^{\prime\prime}},\,
\frac{1}{1-y^{\prime\prime}}
\right),
\end{align}
and the domain and boundaries for $F_5(q)$ in the $u_{1}u_{2}$-plane 
are summarized in Fig.~\ref{fig:F5}.

Using the above results, we can express $F_5(q)$ in
terms of the five-point cross-ratios:
\begin{align}
F_5(q)
&=
\int_0^1dx\int_0^1dy \frac{-i\alpha'}{\pi}
\Big(\frac{q}{2}+p^0_3+
p^0_1\,(1-xy)
+p^0_2\,(1-x)\Big)
\nn
&\quad
\times x^{2\alpha'p_0\cdot p_4+\alpha' p_0\cdot p_0-1}
(1-x)^{2\alpha'p_2\cdot p_4}
y^{2\alpha'p_1\cdot p_3}
(1-y)^{2\alpha'p_1\cdot p_2}
(1-xy)^{2\alpha'p_1\cdot p_4}
\\
&=
\int_0^1dx'\int_0^1dy' \frac{-i\alpha'}{\pi}
\Big(\frac{q}{2}+p^0_3+p^0_1 \,y'
+p^0_2\,x'y'\Big)
\Big)
\nn
&\quad
\times {x'}^{2\alpha'p_2\cdot p_4}
(1-x')^{2\alpha'p_1\cdot p_2}
{y'}^{2\alpha'p_0\cdot p_3+\alpha'p_0\cdot p_0-1}
(1-y')^{2\alpha'p_1\cdot p_3}
(1-x'y')^{2\alpha'p_2\cdot p_3}
\\
&=
\int_0^1dx''\int_0^1dy'' \frac{-i\alpha'}{\pi}
\Big(\frac{q}{2}+p^0_3+
p^0_1\frac{1-y''}{1-x'' y''}
+p^0_2\,(1-y'')\Big)
\Big)
\nn
&\quad
\times {x''}^{2\alpha'p_1\cdot p_2}
(1-x'')^{2\alpha'p_1\cdot p_3}
{y''}^{2\alpha'p_0\cdot p_4+\alpha'p_0\cdot p_0-1}
(1-y'')^{2\alpha'p_0\cdot p_3+\alpha'p_0\cdot p_0-1}
(1-x''y'')^{2\alpha'p_0\cdot p_1}.
\end{align}
We expand the integrand of $F_5(q)$ around the boundaries associated 
with the zeroth vertex (${\cal B}_{40}$ and ${\cal B}_{03}$).
Then, by using the $i\varepsilon$ prescription, 
the singularities are extracted as (\ref{eq:F5delta}).

\subsection{$F_6(q)$}

\begin{figure}[htbp]
\begin{center}
\includegraphics[width=7cm]{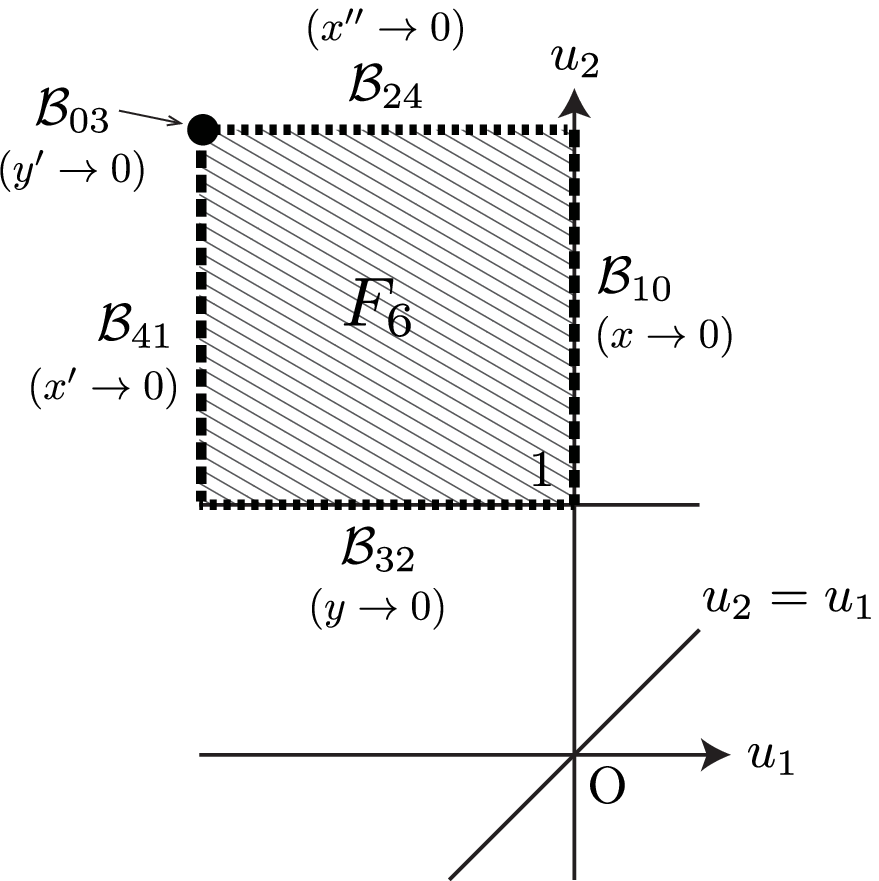}
\end{center}
\caption{The correspondence of the integration domain and the moduli space
for $F_6(q)$.
\label{fig:F6}}
\end{figure}

For $F_6(q)$, we can map the integration region in Fig.~\ref{fig:uv} 
onto the unit square by
\begin{align}
x=\frac{u_{1}}{u_{1}-1},\qquad y=\frac{u_{2}-1}{u_{2}}.
\end{align}
After that, the five-point cross-ratios are introduced 
by (\ref{eq:xytrans}) and (\ref{eq:xytrans2}):
\begin{align}
&x=-\dfrac{y_{01}y_{34}}{y_{04}y_{13}},
&&y=-\dfrac{y_{04}y_{23}}{y_{02}y_{34}},
&&x^{\prime}=\dfrac{y_{02}y_{14}}{y_{04}y_{12}},
&&y^{\prime}=\dfrac{y_{03}y_{12}}{y_{02}y_{13}},
&&x^{\prime\prime}=\dfrac{y_{13}y_{24}}{y_{12}y_{34}}.
\end{align}
Therefore, 
the boundaries of the moduli space are given by
\begin{align}
&{\cal B}_{10}:(x\to 0),
&&{\cal B}_{32}:(y\to 0),
&&{\cal B}_{41}:(x^{\prime}\to 0),
&&{\cal B}_{03}:(y^{\prime}\to 0),
&&{\cal B}_{24}:(x^{\prime\prime}\to 0).
\end{align}
$(u_{1},u_{2})$ is written by
\begin{align}
&(u_1,\,
u_2)=\left(\frac{x}{x-1},\frac{1}{1-y}\right)
=\left(\frac{x^{\prime}y^{\prime}-1}{x^{\prime}y^{\prime}},\,
\frac{1-x^{\prime}y^{\prime}}{y^{\prime}(1-x^{\prime})}\right)
=\left(
\frac{y^{\prime\prime}}{y^{\prime\prime}-1},\,
\frac{1-x^{\prime\prime}y^{\prime\prime}}{x^{\prime\prime}(1-y^{\prime\prime})}
\right),
\end{align}
and the domain and boundaries for $F_6(q)$ in the $u_{1}u_{2}$-plane
 are summarized in Fig.~\ref{fig:F6}.
 
Using the above results, we can express $F_6(q)$ in
terms of the five-point cross-ratios:
\begin{align}
F_6(q)
&=
\int_0^1dx\int_0^1dy \frac{-i\alpha'}{\pi}
\Big(\frac{q}{2}+p^0_3+
p^0_1\frac{x-1}{x}
+p^0_2\,(1-y)\Big)
\nn
&\quad
\times x^{2\alpha'p_0\cdot p_1}
(1-x)^{2\alpha'p_1\cdot p_4}
y^{2\alpha'p_2\cdot p_3}
(1-y)^{2\alpha'p_2\cdot p_4}
(1-xy)^{2\alpha'p_1\cdot p_2}
\\
&=
\int_0^1dx'\int_0^1dy' \frac{-i\alpha'}{\pi}
\Big(\frac{q}{2}+p^0_3+p^0_1\frac{x'y'}{x'y'-1}
+p^0_2\,\frac{(1-x')y'}{1-x'y'}\Big)
\nn
&\quad
\times {x'}^{2\alpha'p_1\cdot p_4}
(1-x')^{2\alpha'p_2\cdot p_4}
{y'}^{2\alpha'p_0\cdot p_3+\alpha'p_0\cdot p_0-1}
(1-y')^{2\alpha'p_2\cdot p_3}
(1-x'y')^{2\alpha'p_3\cdot p_4-\alpha'p_0\cdot p_0+1}
\\
&=
\int_0^1dx''\int_0^1dy'' \frac{-i\alpha'}{\pi}
\Big(\frac{q}{2}+p^0_3+
p^0_1\frac{y''-1}{y''}
+p^0_2\,\frac{x''(1-y'')}{1-x''y''}\Big)
\nn
&\quad
\times {x''}^{2\alpha'p_2\cdot p_4}
(1-x'')^{2\alpha'p_2\cdot p_3}
{y''}^{2\alpha'p_0\cdot p_1}
(1-y'')^{2\alpha'p_0\cdot p_3+\alpha'p_0\cdot p_0-1}
(1-x''y'')^{2\alpha'p_0\cdot p_2}.
\end{align}
We expand the integrand of $F_6(q)$ around the boundaries associated 
with the zeroth vertex (${\cal B}_{10}$ and ${\cal B}_{03}$). 
Then, by using the $i\varepsilon$ prescription, 
the singularities are given as (\ref{eq:F6delta}).



\begin{thebibliography}{99}
\bibitem{Erbin:2019uiz}
H.~Erbin, J.~Maldacena and D.~Skliros,
``Two-Point String Amplitudes,''
JHEP \textbf{07} (2019) 139
[arXiv:1906.06051 [hep-th]].

\bibitem{Seki:2019ycz}
S.~Seki and T.~Takahashi,
``Two-point String Amplitudes Revisited by Operator Formalism,''
Phys. Lett. B \textbf{800} (2020) 135078
[arXiv:1909.03672 [hep-th]].

\bibitem{Seki:2021ivm}
S.~Seki and T.~Takahashi,
``Reduction of Open String Amplitudes by Mostly BRST Exact Operators,''
[arXiv:2108.05628 [hep-th]].

\bibitem{Kashyap:2020tgx}
S.~P.~Kashyap,
``Two-Point Superstring Tree Amplitudes Using the Pure Spinor Formalism,''
[arXiv:2012.03802 [hep-th]].

\bibitem{Witten:2013pra}
E.~Witten,
``The Feynman $i \epsilon$ in String Theory,''
JHEP \textbf{04} (2015) 055
[arXiv:1307.5124 [hep-th]].

\bibitem{Veneziano:1968yb}
G.~Veneziano,
``Construction of a crossing - symmetric, Regge behaved amplitude for linearly rising trajectories,''
Nuovo Cim. A \textbf{57} (1968) 190-197.

\bibitem{Bardakci:1968rse}
K.~Bardakci and H.~Ruegg,
``Reggeized resonance model for the production amplitude,''
Phys. Lett. B \textbf{28} (1968) 342-347.

\bibitem{Virasoro:1969pd}
M.~A.~Virasoro,
``Generalization of veneziano's formula for the five-point function,''
Phys. Rev. Lett. \textbf{22} (1969) 37-39.

\bibitem{Koba:1969rw}
Z.~Koba and H.~B.~Nielsen,
``Reaction amplitude for n mesons: A Generalization of the Veneziano-Bardakci-Ruegg-Virasora model,''
Nucl. Phys. B \textbf{10} (1969) 633-655.

\bibitem{Hanson_Sha}
A.~J.~Hanson and J.~Sha,
``A Contour Integral Representation for the Dual Five-Point Function and a
Symmetry of the Genus Four Surface in R6,''
J.~Phys.~A: Math.~Gen. 39 (2006) 2509-2537
[arXiv:math-ph/0510064].

\bibitem{Polchinski:1998rq}
J.~Polchinski,
{\it String theory. Vol. 1: An introduction to the bosonic string},
Cambridge Univ. Pr., UK, 1998.


\end{thebibliography}
\end{document}